\documentclass[preprint,superscriptaddress,preprintnumbers,showpacs]{revtex4}
\usepackage{amssymb}
\usepackage{graphicx}
\usepackage{dcolumn}
\usepackage{bm}

\begin{document}

\newcommand*{\PKU}{School of Physics and State Key Laboratory of Nuclear Physics and
Technology, Peking University, Beijing 100871}\affiliation{\PKU}
\newcommand*{\NTU}{Department of Physics and Center for Theoretical Sciences, National Taiwan University, Taipei 10617}\affiliation{\NTU}
\newcommand*{\KITPC}{Kavli Institute for Theoretical Physics China, Chinese Academy of Sciences, Beijing 100190}\affiliation{\KITPC}

\title{Triminimal Parametrization of Quark Mixing Matrix}

\author{Xiao-Gang He}\affiliation{\PKU}\affiliation{\NTU}\affiliation{\KITPC}
\author{Shi-Wen Li}\affiliation{\PKU}
\author{Bo-Qiang Ma}\email{mabq@phy.pku.edu.cn}\affiliation{\PKU}\affiliation{\KITPC}

\begin{abstract}
Starting from a new zeroth order basis for quark mixing (CKM) matrix
based on the quark-lepton complementarity and the tri-bimaximal
pattern of lepton mixing, we derive a triminimal parametrization of
CKM matrix with three small angles and a CP-violating phase as its
parameters. This new triminimal parametrization has the merits of
fast convergence and simplicity in application. With the
quark-lepton complementary relations, we derive relations between
the two unified triminimal parametrizations for quark mixing
obtained in this work and for lepton mixing obtained by
Pakvasa-Rodejohann-Weiler. Parametrization deviating from
quark-lepton complementarity is also discussed.

\end{abstract}


\maketitle

Quarks are the fundamental constituents of matter. An appropriate
parametrization of the quark mixing matrix is an important step to
understand the mixing pattern and its underlying theory. The mixing
matrix can be described by the Cabibbo~\cite{cabibbo}, Kobayashi and
Maskawa~\cite{km}(CKM) $V_{\mathrm{CKM}}$ matrix. A commonly used
form of $V_{\mathrm{CKM}}$ for three generations of quarks is given
by~\cite{ck,yao},
\begin{eqnarray}{\small \left(
\begin{array}{ccc}
c_{12}c_{13} & s_{12}c_{13} & s_{13}e^{-i\delta}           \\
-s_{12}c_{23}-c_{12}s_{23}s_{13}e^{i\delta} &
c_{12}c_{23}-s_{12}s_{23}s_{13}e^{i\delta}  & s_{23}c_{13} \\
s_{12}s_{23}-c_{12}c_{23}s_{13}e^{i\delta}  &
-c_{12}s_{23}-s_{12}c_{23}s_{13}e^{i\delta} & c_{23}c_{13}
\end{array}
\right)},\label{ckmp}
\end{eqnarray}
where $s_{ij}=\sin\theta_{ij}$ and $c_{ij}=\cos\theta_{ij}$ are the
rotation angles and $\delta$ is the CP violating phase. The above
expression can be decomposed as
\begin{eqnarray}
V_{\mathrm{CKM}}&=&R_{23}(\theta_{23})U_\delta^\dag
R_{13}(\theta_{13})U_\delta R_{12}(\theta_{12})\;,
\end{eqnarray}
with
\begin{eqnarray}
&&R_{23}=\left(
\begin{array}{ccc}
1 & 0 & 0           \\
0 & c_{23} & s_{23} \\
0 & -s_{23} & c_{23}\\
\end{array}
\right)\;,\;\;R_{13}=\left(
\begin{array}{ccc}
c_{13} & 0 & s_{13}           \\
0 & 1 & 0 \\
-s_{13} & 0 & c_{13}\\
\end{array}
\right)\;, \quad R_{12} = \left(
\begin{array}{ccc}
c_{12}& s_{12} & 0           \\
-s_{12} & c_{12} & 0 \\
0 & 0 & 1\\
\end{array}
\right)\;,\label{ex}
\end{eqnarray}
and $U_{\delta}$ is a diagonal phase matrix with the diagonal
elements given by ${\mathrm{diag}}(e^{i\delta/2}, 1,
e^{-i\delta/2})$.

The elements in the CKM matrix have been determined to high
precision with the allowed ranges of the magnitudes of its elements
as~\cite{yao}
%\begin{widetext}
\begin{eqnarray}{\small \left(
  \begin{array}{ccc}
    0.97419\pm0.00022             & 0.2257\pm0.0010    & 0.00359\pm0.00016               \\
    0.2256\pm0.0010               & 0.97334\pm0.00023  & 0.0415^{+0.0010}_{-0.0011}      \\
    0.00874^{+0.00026}_{-0.00037} & 0.0407\pm0.0010    & 0.999133^{+0.000044}_{-0.000043}
  \end{array} \right)}.\label{vv}
\end{eqnarray}
%\end{widetext}
With Eq.~(\ref{ckmp}) and the data given in Eq.~(\ref{vv}), we find
that
\begin{eqnarray}
&&\sin\theta_{12}=\frac{|V_{us}|}{\sqrt{|V_{ud}|^2+|V_{us}|^2}}=0.2257\pm0.0010,\nonumber\\
&&\sin\theta_{23}=\frac{|V_{cb}|}{\sqrt{|V_{ud}|^2+|V_{us}|^2}}=0.0415^{+0.0010}_{-0.0011},\nonumber\\
&&\sin\theta_{13}=|V_{ub}|=0.00359\pm0.00016.\label{sn}
\end{eqnarray}
The CP violating phase has also been determined. A recent full fit
from UTfit group gives $\delta=(66.7\pm 6.4)^\circ$~\cite{utfit}.

The CKM matrix is close to the unit matrix with small deviations in
the non-diagonal elements. This led Wolfenstein, some time ago, to
parameterize the mixing matrix as~\cite{wolfensteinpara},
\begin{eqnarray}{\small V_{\mathrm{CKM}}=\left(
  \begin{array}{ccc}
    1-\frac{1}{2}\lambda^2   & \lambda                & A\lambda^3(\rho-i\eta) \\
    -\lambda                 & 1-\frac{1}{2}\lambda^2 & A\lambda^2             \\
    A\lambda^3(1-\rho-i\eta) & -A\lambda^2            & 1                      \\
  \end{array}
\right)+\mathcal{O}(\lambda^4)}.\label{wolfenstein}
\end{eqnarray}
This parametrization explicitly shows the deviations of the
non-diagonal elements from the unit matrix in different powers of
the parameter $\lambda$ with $\lambda=0.2257^{+0.0009}_{-0.0010}$.
Since $A=0.814^{+0.021}_{-0.022}$, $\rho(1-\lambda^2/2+\dots) =
0.135^{+0.031}_{-0.016}$, and $\eta(1-\lambda^2/2+\dots) =
0.349^{+0.015}_{-0.017}$~\cite{yao} which are of order one, the
hierarchy of the CKM elements in powers of $\lambda$ in the
expansion is evident.

The unit matrix as the zeroth leading order for expanding the CKM
matrix is easy to use. However, we find that parameterizing the CKM
matrix in ways leading to faster convergence also interesting which
provides more information about mixing. One then asks the question that what
physical reason(s) would help to make the zeroth leading order matrix.
An appealing guideline in
searching for such a method is the quark-lepton complementarity
(QLC)~\cite{smirnov,qlc}, first introduced by Smirnov~\cite{smirnov}
and then discussed in many other papers~\cite{qlc}, which relates
quark and lepton mixing angles with
\begin{eqnarray}
\theta_{12}^Q+\theta_{12}^L=\frac{\pi}{4},\quad
\theta_{23}^Q+\theta_{23}^L=\frac{\pi}{4},\quad
\theta_{13}^Q\sim\theta_{13}^L\sim0,\label{qlc}
\end{eqnarray}
where the superscript $Q$ indicates the mixing angles in the CKM
matrix, and the superscript $L$ indicates mixing angles in the
lepton mixing matrix $U_{\mathrm{PMNS}}$, i.e, the
Pontecorvo~\cite{pontecorvo}-Maki-Nakawaga-Sakata~\cite{mns} (PMNS)
matrix. From now on we will use these notations to distinguish
mixing angles in the CKM and PMNS matrices. The above relations
enable one to determine the mixing angles use both quark and lepton
mixing information, and therefore provide a unified understanding of
quark mixing and lepton mixing phenomena.

Current experimental data show that the PMNS matrix is close to the
tri-bimaximal pattern given by,
\begin{eqnarray}
    U=\left(
        \begin{array}{ccc}
            2/\sqrt{6}  & 1/\sqrt{3}  & 0          \\
           -1/\sqrt{6}  & 1/\sqrt{3}  & 1/\sqrt{2} \\
            1/\sqrt{6}  & -1/\sqrt{3} & 1/\sqrt{2}
        \end{array}
        \right).\label{utribi}
\end{eqnarray}

The tri-bimaximal mixing has been studied extensively in
literature~\cite{harrison, xing}. Parametrization of the PMNS matrix
around the tri-bimaximal pattern has been treated in several
approaches~\cite{linan}. Recently, Pakvasa, Rodejohann, and Weiler
(PRW) gave a new parametrization of the PMNS matrix~\cite{pakvasa}
with the matrix in Eq.~(\ref{utribi}) as the zeroth order in an
expansion. In their paper, the parameters chosen are not the
traditional deviations within the elements of the matrix, instead,
they are the deviations from the tri-bimaximal angles. The method
pointed out a new way to parameterize the mixing matrix with all
angles small, i.e. the ``triminimal'' parametrization.

Enlightened by above idea, we propose to parameterize the CKM matrix
with also three small angles, which we refer as ``triminimal''
parametrization of the CKM matrix.  In this parametrization the
zeroth leading order CKM matrix $V_b$ is inspired from QLC suggested
in Ref.~\cite{li}, \vspace{-0.3cm}
\begin{eqnarray}
\sin\theta^Q_{12}=\frac{\sqrt{2}-1}{\sqrt{6}},\quad\sin\theta^Q_{23}=0,\quad\sin\theta^Q_{13}=0.\label{vbs}
\end{eqnarray}
which leads to the zeroth order basis as
\begin{eqnarray}
    V_b=\left(
        \begin{array}{ccc}
            \frac{\sqrt{2}+1}{\sqrt{6}} & \frac{\sqrt{2}-1}{\sqrt{6}} & 0 \\
           -\frac{\sqrt{2}-1}{\sqrt{6}} & \frac{\sqrt{2}+1}{\sqrt{6}} & 0 \\
            0                           & 0                           & 1
        \end{array}
        \right).\label{vb}
\end{eqnarray}

Although the new matrix in Eq.~(\ref{vb}) is a little more
complicated compared with the unit matrix, it is closer to reality
and deviations of the CKM matrix from the new matrix are rather
small. The expansion around $V_b$ converges very quickly.

We have motivated our choice of $V_b$ from QLC relations following
Ref.~\cite{li}. If a different principle is used, the form $V_b$ may
be different. For example, it has been argued that the angle
$\theta_{12}^Q$ should satisfy~\cite{golden} $\tan(2\theta_{12}) =
1/2$ which is numerically more close to the experimental value than
we choose. There are ambiguities in choosing the zeroth order
matrix. Better theoretical understanding is needed to determine the
form of $V_b$. We believe that the QLC, which leads to some unified
understanding of mixing angles in lepton and quark sectors, is a
good starting point and will follow this rout to carry out our
analysis. When using QLC relations, one should also be careful about
at which scale they hold. In this work we will assume that QLC holds
at a relatively low scale such that the renormalization group
equation (RGE) running effects down to SM scale can be neglected. If
the QLC holds at a very high energy, say grand unification scale,
one should include possible large RGE running effects, in particular
possible large effects in neutrino sector~\cite{Babu}. This is an
interesting possibility which will be studied in a future work.

Using $\epsilon^Q_{12}$, $\epsilon^Q_{23}$, $\epsilon^Q_{13}$
defined by \vspace{-0.2cm}
\begin{eqnarray}
\theta^Q_{12}=\arcsin\frac{\sqrt{2}-1}{\sqrt{6}}+\epsilon^Q_{12},
\quad\theta^Q_{23}=\epsilon^Q_{23},\quad\theta^Q_{13}=\epsilon^Q_{13},\label{set}
\end{eqnarray}
we obtain the triminimal form for the CKM matrix, \vspace{-0.3cm}
\begin{eqnarray}
V_{\mathrm{CKM}}&=&R_{23}(\theta^Q_{23})U^\dag_\delta
R_{13}(\theta^Q_{13})U_\delta
R_{12}(\theta^Q_{12})\nonumber\\
&=&R_{23}(\epsilon^Q_{23})U^\dag_\delta
R_{13}(\epsilon^Q_{13})U_\delta
R_{12}(\epsilon^Q_{12})R_{12}(\arcsin\frac{\sqrt{2}-1}{\sqrt{6}})\nonumber\\
&=&R_{23}(\epsilon^Q_{23})U^\dag_\delta
R_{13}(\epsilon^Q_{13})U_\delta R_{12}(\epsilon^Q_{12})V_b.
\end{eqnarray}
Comparing with the data in Eq.~(\ref{sn}), we obtain
\begin{eqnarray}
&&\epsilon^Q_{12}=0.0577\pm0.0010,\quad\epsilon^Q_{23}=0.0415^{+0.0010}_{-0.0011},\nonumber\\
&&\epsilon^Q_{13}=0.00359\pm0.00016.\label{vn}
\end{eqnarray}
We see that the angles $\epsilon^Q_{ij}$ are indeed very small, much
smaller than $\lambda$ in the Wolfenstein parametrization. Therefore
using the parameters $\epsilon^Q_{ij}$, the expansion converges
faster than the Wolfenstein parametrization.

\begin{table*}
\caption{\label{tab} Values of some parameters based on the global
analysis of lepton mixing and quark mixing data. The first and
second errors are for $1\sigma$ and $3\sigma$ error ranges.}
\begin{ruledtabular}
{\scriptsize
\begin{tabular}{ccccccc}
$\theta_{12}^Q+\theta_{12}^L$ & $\theta_{23}^Q+\theta_{23}^L$ &
$\epsilon^L_{12}$ & $\epsilon^L_{23}$ &
$\epsilon^Q_{12}+\epsilon^L_{12}$ & $\epsilon^Q_{23}+\epsilon^L_{23}$ & literature \\
\hline
$47.5^{\circ+1.4^\circ}_{\,\,\,-1.4^\circ}(^{+4.8^\circ}_{-4.0^\circ})$
&
$44.7^{\circ+5.1^\circ}_{\,\,\,-3.3^\circ}(^{+11.3^\circ}_{-7.7^\circ})$
& $-0.01^{+0.02}_{-0.02}(^{+0.08}_{-0.07})$ &
$-0.05^{+0.09}_{-0.06}(^{+0.20}_{-0.13})$ &
$0.04^{+0.02}_{-0.02}(^{+0.08}_{-0.07})$ &
$0.01^{+0.09}_{-0.06}(^{+0.20}_{-0.13})$ & Ref.~\cite{gonzalez}  \\
$47.0^{\circ+1.2^\circ}_{\,\,\,-1.1^\circ}(^{+3.9^\circ}_{-3.0^\circ})$
&
$45.4^{\circ+4.2^\circ}_{\,\,\,-3.3^\circ}(^{+10.2^\circ}_{-7.8^\circ})$
& $-0.02^{+0.02}_{-0.02}(^{+0.07}_{-0.05})$ &
$-0.03^{+0.07}_{-0.06}(^{+0.18}_{-0.14})$ &
$0.03^{+0.02}_{-0.02}(^{+0.07}_{-0.05})$ &
$0.01^{+0.07}_{-0.06}(^{+0.18}_{-0.14})$ & Ref.~\cite{fogli}    \\
$46.5^{\circ+1.4^\circ}_{\,\,\,-1.0^\circ}(^{+4.4^\circ}_{-3.1^\circ})$
&
$47.4^{\circ+4.0^\circ}_{\,\,\,-3.4^\circ}(^{+9.7^\circ}_{-8.0^\circ})$
& $-0.03^{+0.02}_{-0.02}(^{+0.08}_{-0.05})$ &
~~$0.00^{+0.07}_{-0.06}(^{+0.17}_{-0.14})$ &
$0.03^{+0.02}_{-0.02}(^{+0.08}_{-0.05})$ &
$0.04^{+0.07}_{-0.06}(^{+0.17}_{-0.14})$ & Ref.~\cite{schwetz}   \\
\end{tabular}}
\end{ruledtabular}
\end{table*}

To the second order expansion in $\epsilon^Q_{ij}$,
$V_{\mathrm{CKM}}$ is given by
\begin{widetext}
\vspace{-0.7cm} {\footnotesize
\begin{eqnarray}
V_{\mathrm{CKM}} &=&V_b
+ \epsilon^Q_{12}\left(\begin{array}{ccc}   -\frac{\sqrt{2}-1}{\sqrt{6}} &  \frac{\sqrt{2}+1}{\sqrt{6}} & 0 \\
                                           -\frac{\sqrt{2}+1}{\sqrt{6}} & -\frac{\sqrt{2}-1}{\sqrt{6}} & 0 \\
                                           0 & 0 & 0 \\
                                           \end{array}
                                           \right)
  +\epsilon^Q_{23}\left(
                  \begin{array}{ccc}
                    0 & 0 & 0 \\
                    0 & 0 & 1 \\
                    \frac{\sqrt{2}-1}{\sqrt{6}} & -\frac{\sqrt{2}+1}{\sqrt{6}} & 0 \\
                  \end{array}
                \right)
  +\epsilon^Q_{13}\left(
                  \begin{array}{ccc}
                    0 & 0 & e^{-i\delta} \\
                    0 & 0 & 0 \\
                    -\frac{\sqrt{2}+1}{\sqrt{6}}e^{i\delta} & -\frac{\sqrt{2}-1}{\sqrt{6}}e^{i\delta} & 0 \\
                  \end{array}
                \right)\nonumber\\
  &+&(\epsilon^Q_{12})^2\left(
                  \begin{array}{ccc}
                    -\frac{\sqrt{2}+1}{2\sqrt{6}} & -\frac{\sqrt{2}-1}{2\sqrt{6}} & 0 \\
                    \frac{\sqrt{2}-1}{2\sqrt{6}} & -\frac{\sqrt{2}+1}{2\sqrt{6}} & 0 \\
                    0 & 0 & 0 \\
                  \end{array}
                \right)
  +(\epsilon^Q_{23})^2\left(
                  \begin{array}{ccc}
                    0 & 0 & 0 \\
                    \frac{\sqrt{2}-1}{2\sqrt{6}} & -\frac{\sqrt{2}+1}{2\sqrt{6}} & 0 \\
                    0 & 0 & -\frac{1}{2} \\
                  \end{array}
                \right)
  +(\epsilon^Q_{13})^2\left(
                  \begin{array}{ccc}
                    -\frac{\sqrt{2}+1}{2\sqrt{6}} & -\frac{\sqrt{2}-1}{2\sqrt{6}} & 0 \\
                    0 & 0 & 0 \\
                    0 & 0 & -\frac{1}{2} \\
                  \end{array}
                \right)\nonumber\\
  &+&\epsilon^Q_{12}\epsilon^Q_{23}\left(
                  \begin{array}{ccc}
                    0 & 0 & 0 \\
                    0 & 0 & 0 \\
                    \frac{\sqrt{2}+1}{\sqrt{6}} & \frac{\sqrt{2}-1}{\sqrt{6}} & 0 \\
                  \end{array}
                \right)
  +\epsilon^Q_{12}\epsilon^Q_{13}e^{i\delta}\left(
                  \begin{array}{ccc}
                    0 & 0 & 0 \\
                    0 & 0 & 0 \\
                    \frac{\sqrt{2}-1}{\sqrt{6}} & -\frac{\sqrt{2}+1}{\sqrt{6}} & 0 \\
                  \end{array}
                \right)
  +\epsilon^Q_{23}\epsilon^Q_{13}e^{i\delta}\left(
                  \begin{array}{ccc}
                    0 & 0 & 0 \\
                    -\frac{\sqrt{2}+1}{\sqrt{6}} & -\frac{\sqrt{2}-1}{\sqrt{6}} & 0 \\
                    0 & 0 & 0 \\
                  \end{array}
                \right).\label{ve}
\end{eqnarray}}
\end{widetext}

Note that numerically we have, $\epsilon^Q_{23}\sim\epsilon^Q_{12}$,
but $\epsilon^Q_{13}\sim (\epsilon^Q_{12})^2$. This implies that
$\epsilon^Q_{12}$, $\epsilon^Q_{23}$ are of order $\mathcal
{O}(\epsilon^Q_{12})$, but $\epsilon^Q_{13}$ is of order $\mathcal
{O}((\epsilon^Q_{12})^2)$. Therefore
$\epsilon^Q_{12}\epsilon^Q_{13}$, $\epsilon^Q_{23}\epsilon^Q_{13}$
and $(\epsilon^Q_{13})^2$ in Eq.~(\ref{ve}) can be omitted if we
make an approximation to order $\mathcal {O}((\epsilon^Q_{12})^2)$. Also
note that the effect of CP violation exists in first power in
$\epsilon^Q_{13}$, it is of order $\mathcal {O}((\epsilon^Q_{12})^2)$ in
this parametrization, which is coordinated with the result in
Ref.~\cite{li}.

The rephasing-invariant Jarlskog parameter, $\mathcal{J}\equiv{\rm
Im}(V_{us}V_{cb}V_{ub}^\ast V_{cs}^\ast)$~\cite{jarlskog}, in our
parametrization is given by a simple expression,
\begin{eqnarray}
\mathcal{J}=\left(\frac{1}{6}+\frac{2\sqrt{2}}{3}\epsilon^Q_{12}\right)
\epsilon^Q_{23}\epsilon^Q_{13}\sin\delta.
\end{eqnarray}
Since
$\frac{2\sqrt{2}}{3}\epsilon^Q_{12}\sim\frac{1}{3}\times\frac{1}{6}$,
the $\mathcal {O}((\epsilon^Q_{ij})^3)$ term should be kept in the
expression of $\mathcal{J}$.

The QLC relations in Eq.~(\ref{qlc}) has important implications. If
one imposes these empirical equations, the parameters in our
parametrization and those proposed for lepton mixing in
Ref.~\cite{pakvasa} are not independent, we find that
\begin{eqnarray}
&&\epsilon_{12}^Q+\epsilon_{12}^L=0,\nonumber\\
&&\epsilon_{23}^Q+\epsilon_{23}^L=0,\nonumber\\
&&\epsilon_{13}^Q\sim\epsilon_{13}^L\sim0. \label{newqlc}
\end{eqnarray}
The QLC leads to unified triminimal parametrizations for mixing in
quark and lepton sectors.

Considerable experimental data on lepton mixing have been
accumulated. To have some idea how good QLC relations provide
guidance, in Table \ref{tab} we show the values of several
quantities with their $1\sigma(3\sigma)$ errors, for QLC relations using global data
fitting on CKM matrix elements in Eq.~(\ref{vv})~\cite{yao}, and on
neutrino mixing angles obtained in~\cite{gonzalez}, \cite{fogli} and
\cite{schwetz}.

We see, from Table \ref{tab}, that there are differences from
different fits which reflects possible theoretical uncertainties in
handling row data. However all fits obtain similar results. The best
fit values of $\epsilon_{12, 23}^L$ are minus relative to
$\epsilon^Q_{12,23}$ which are in the right direction. But there are
deviations. The values for $\epsilon^L_{12, 23}$ are different from
$-\epsilon^Q_{12,23}$. In particular the value for $\epsilon^L_{12}$
is substantially away from $\epsilon^Q_{12}$. The QLC relation for
$\epsilon_{23}^{Q,L}$ is satisfied at $1\sigma$ level. The QLC
relation for $\epsilon^{Q,L}_{12}$ is inconsistence at 1$\sigma$
level. There is a tension with QLC predictions. However all QLC
predictions are consistent at $3\sigma$ level. More experimental
data are needed to test QLC. If the current trend persists, one
should modify the QLC to some extent.

In most applications of QLC, $\epsilon_{13}^Q$ and $\epsilon_{13}^L$
do not have a specific relation, what is known is that they are both
small. In this case the well determined $\epsilon^Q_{13}$ will not
provide severe constraint on $\epsilon^L_{13}$. If one requires
$\epsilon^Q_{13} + \epsilon^L_{13} =0$ or $\epsilon^L_{13} =
\epsilon^Q_{13} $, then the value for $U_{e3}$ will be too small to
be measured at near future experiments. However, this needs not be
true as $\epsilon^Q_{12}$ and $\epsilon^L_{12}$ can be independent.
At present it is not known whether the CP violating phases
$\delta^Q$ and $\delta^L$ for CKM and PMNS matrices are related.
Therefore knowing $\delta^Q$, one still does not know much about
$\delta^L$. The determinations of $\epsilon^L_{13}$ and $\delta^L$
have to be left to future experiments or a underlying theory for CKM
and PMNS matrices to decide. Also note that no information about
possible Majorana phases in the PMNS matrix can be inferred from QLC
in the present form.

We would like to stress that the triminimal parametrizations do not
depend on QLC assumption since our parametrization and the parametrization in
Ref.~\cite{pakvasa} are both general and can be independent of each
other. If future experiments will rule out the QLC relations, one
should not impose the relations in Eq.~(\ref{qlc}). The deviations
from QLC relations, to the second order in $\epsilon^{Q,L}_{ij}$,
can be written as
\begin{eqnarray}
&&\sin(\theta_{12}^Q+\theta_{12}^L)=\frac{1}{\sqrt{2}}\left(\sin(\epsilon^Q_{12}+
\epsilon^L_{12})
+ \cos(\epsilon^Q_{12}+\epsilon^L_{12})\right)\nonumber\\
&&\approx \frac{1}{\sqrt{2}}\left(
1+(\epsilon_{12}^Q+\epsilon_{12}^L)
-\frac{1}{2}(\epsilon_{12}^Q+\epsilon_{12}^L)^2\right),\nonumber\\
&&\sin(\theta_{23}^Q+\theta_{23}^L)=
\frac{1}{\sqrt{2}}\left(\sin(\epsilon^Q_{23}+ \epsilon^L_{23})
+ \cos(\epsilon^Q_{23}+\epsilon^L_{23})\right)\nonumber\\
&&\approx \frac{1}{\sqrt{2}}\left(1
+(\epsilon_{23}^Q+\epsilon_{23}^L)
-\frac{1}{2}(\epsilon_{23}^Q+\epsilon_{23}^L)^2\right).
\end{eqnarray}
Clearly, the QLC is satisfied to zeroth order. That is to say, the
QLC may be considered to be the lowest order approximation of the
sum of the corresponding mixing angles of the CKM and PMNS matrices.

As pointed out earlier, there are evidences of violation of QLC
since the best fit values
$\epsilon^Q_{12}+\epsilon^L_{12}=(0.04,\;0.03,\;0.03)$ and
$\epsilon^Q_{23}+\epsilon^L_{23}=(0.01,\;0.01,\;0.04)$ shown in
Table \ref{tab} deviate from zero. One has to wait more data to make
conclusions. In any case, parametrization deviating from QLC given
above can provide a convenient way for experimental studies.

To conclude, we have proposed a new triminimal parametrization of
the CKM matrix starting from a new zeroth order basis of quark
mixing based on the quark-lepton complementarity and the
tri-bimaximal pattern of lepton mixing. This new parametrization has
three small angles and a CP-violating phase as its parameters. It
not only has a unified form similar to the recently known PRW
triminimal parametrization of lepton mixing~\cite{pakvasa}, but also
has the merits of fast convergence and simplicity in application.
With the quark-lepton complementary relations, we have derived relations
between the two unified triminimal parametrizations for both quark
mixing and lepton mixing. Parametrization deviating from
quark-lepton complementarity is also discussed.

\begin{acknowledgments}
This work is partially supported by NSC, NCTS, NSFC (Nos.~10721063,
1057-5003, 10528510), by the Key Grant Project of Chinese Ministry
of Education (No.~305001), and by the Research Fund for the Doctoral
Program of Higher Education.
\end{acknowledgments}

\end{document}